\documentclass[referee]{aa} 
\usepackage{epsfig}
\def\beq{\begin{equation}}
\def\eeq{\end{equation}}
\begin{document}

\title{Fast Magnetosonic Waves Driven by Gravitational Waves}

\author{D. Papadopoulos \inst{1}, N. Stergioulas \inst{1},
  L. Vlahos \inst{1}
  \and
  J.~Kuijpers \inst{2,3}}
\institute{Department of Physics, Aristoteleion University of
Thessaloniki, \\54006 Thessaloniki, Greece 
\and Department of Astrophysics,
University of Nijmegen, PO Box 9010, \\6500
  GL Nijmegen, The Netherlands
  \and
Astronomical Institute, Utrecht University, PO Box 80 000, \\3508
TA Utrecht, The Netherlands}

\offprints{D. Papadopoulos}

\date{Received .........; accepted .............}
\titlerunning{Magnetosonic waves driven by Gravitational waves}
\authorrunning{D. Papadopoulos et al....}

\maketitle

\begin{abstract} The propagation of a gravitational wave (GW) through a
  magnetized plasma is considered. In particular, we study the
  excitation of fast magnetosonic waves (MSW) by a gravitational wave,
  using the linearized general-relativistic hydromagnetic equations.
  We derive the dispersion relation for the plasma, treating
  the gravitational wave as a perturbation in a Minkowski
  background space-time. We show that the presence  of
  gravitational waves will drive magnetosonic waves in the plasma
  and discuss the potential astrophysical implications.
\end{abstract}
   \keywords{gravitational
    waves--MHD waves--compact objects: pulsars; magnetars; GRBs.}

\section{Introduction}
It is well known that gravitational waves (GWs) can interact with
electromagnetic waves.
The scattering of electromagnetic waves by time dependent
GWs has also been considered by several authors
(DeWitt \& Breheme 1960; Cooperstock 1968; Chestrs, 1973; Denisov
1978; Grischuk \& Polnarev 1980; Macdonald \& Thorne 1982; Machedo \& Nelson 1983;
Demianski 1983; Marklund et al. 2000). It has been shown that the
coupling is mainly due to the generation of electric currents
inside the plasma by the perturbation of the charged particle
trajectories by the GW.
Several studies used the kinetic theory formulation to study the
coupling of GWs with electromagnetic waves in a
magnetized vacuum (Gertsenshtein 1962; Zeldovich 1974) or near a
charged black hole (Gerlach 1974). The first attempt to study the
propagation of GWs in a magnetized plasma using the
kinetic theory was by Macedo and Nelson (1983). They studied the
coupling of GWs propagating parallel and
perpendicular to the external magnetic field with high-frequency
electromagnetic modes. The coupling of the gravitational with the
high-frequency electromagnetic modes propagating along the
magnetic field was weak. This is due to the fact that the
GW propagating along the direction of the
undisturbed magnetic field does not generate currents in the
plasma. The coupling is substantially higher for ordinary and
extraordinary waves propagating perpendicular to the magnetic
field but the energy transfer of gravitational energy to
electromagnetic energy in the interstellar space is extremely
weak.

Brodin \& Marklund (1999) discuss the parametric excitation of plasma
waves in the presence of gravitational radiation. They assumed that
the GW was a monochromatic wave and estimated, using
Einstein-Maxwell-Vlasov's equations, the growth of Langmuir and
electromagnetic waves in a cold unmagnetized plasma. Marklund et
al. (2000) used the 1+3 orthonormal frame formalism (Ellis \& van
Elst, 1998) to study the linear interaction of a GW inside a
inhomogeneous magnetized plasma. They showed that, close to a strong
gravitational source, the electric fields associated with the
electrostatic waves can reach appreciable levels. The direct escape of
plasma radiation is not possible without a non-linear conversion (see
Brodin et al. 2000). Recently Servin et al.  (2000) discuss the
non-linear parametric excitation of Alfv{\'e}n waves by a strong
monochromatic GW. A nonlinear study of shock waves in a highly
magnetized plasma, driven by GWs has been presented by Ignat'ev (1997).

The non-linear interaction of a monochromatic GW with charged
particles and their cyclotron damping inside a plasma has also been
analyzed in detail (see Esposito 1971; Polnarev 1972; Asseo et
al. 1976; Papadopoulos \& Esposito 1985; Varvoglis and Papadopoulos
1992; Servin et al. 2001)

Papadopoulos and Esposito (1982) derived the exact equations for a
finite amplitude MHD wave propagating inside the plasma, within the
framework of general relativity.

In this article we study, for the first time, the excitation of
magnetosonic waves from a GW propagating inside a uniform, strongly
magnetized plasma (where the Alfv\'en speed approaches the speed of
light). In section 2 we outline the basic equations, in section 3 we
derive the general dispersion relation for the coupling of the GWs
with the MHD waves and in section 4 the dispersion relation for the
coupling of the gravitational with ambient magnetosonic waves. In
section 5 we discuss our results and their potential astrophysical
implications, while section 6 summarizes our conclusions.

\section{Basic equations}

The equations for finite amplitude wave propagation in an ideal
hydromagnetic plasma in the general relativistic framework are (see
Papadopoulos \& Esposito, 1982)

\begin{eqnarray}
(\epsilon&-&\frac{H^2}{2})_{;ab} u^a u^b =
h^{ab}(p+\frac{H^2}{2})_{;ab}+2(H^2\Theta)^.\nonumber\\&-&(H^aH^b)_{;ab}
+2x(\frac{2\Theta^2}{3}+\sigma^2-\omega^2
\nonumber\\&-&\dot{u}^a\dot{u}_{a})+
\frac{x}{2}(\rho+3p\nonumber\\&+&H^2)+2\dot{u}_{a}
(H^aH^b)_{;b}+(H^2)_{;a}\dot{u}^a. \label{1}
\end{eqnarray}
Maxwell's equations take on the compact form \beq
\dot{H}^a=(\sigma_b^a+\omega_b^a-\frac{2}{3}\delta_b^a\Theta )H^b+
\frac{1}{\epsilon+p}p_{;b}H^bu^a, \label{2} \eeq and the equation of
motion is \beq  \dot{x} u^a+x\dot{u}^a+x\Theta
u^a+(p+\frac{H^2}{2})_{;b}g^{ab}-(H^a H^b)_{;b}=0 \eeq where $u^a$ is
the fluid velocity, $\epsilon$ is the energy density, $\rho$ the rest
mass density, $p$ the pressure of the fluid, $H^a=(0,H^1,H^2,H^3)$ is
the prevailing magnetic field, $\dot{u}^a=u_{;c}^a u^c$,
$x=\epsilon+p+H^2,$ $h^{ab}=g^{ab}+u^a u^b,$ $\sigma_{ij}$ is the
shear, $\omega_{ij}$ is an antisymmetric tensor, corresponding to the
rotation of the fluid and $\Theta$ is the expansion of the fluid. In
the equations we have assumed $c=1$, $\kappa=8\pi G/c^4=1$ and we have
put $H=B \to \sqrt{4\pi}H$ (from Gaussian units), where $\kappa$ is
the coupling constant in Einstein's field equations.

We perturb the equations (1)-(2) and keep only linear terms in the
perturbed quantities (see Eqs. (47)-(49) in Papadopoulos and Esposito,
1982), and use the metric given by $g_{\mu \nu}=\eta_{\mu \nu}+h_{\mu
\nu}$ or

\beq ds^2=-c^2 dt^2+(1+h) dx^2+(1-h) dy^2 +dz^2 \eeq where $ h=h(t,z)$
represents the perturbation of the flat space by the GW.

In the following section we will search for the coupling of the normal
modes of the magnetized plasma with the GW. It is important to notice
that both gravitational and MHD waves follow similar dispersion
relations, in the asymptotic limit when the relativistic Alfv{\'e}n
speed  approaches the speed of light (see also Landau and Lifshitz,
1975).

\section{ The Perturbed Equations}

In this section we will perturb Eqs.(1)-(2) assuming  that $\delta
g_{ab}\neq 0$.  Perturbing Eq.(1) we find

\begin{eqnarray} & &\delta[(\epsilon-\frac{H^2}{2})_{;ab}]u^a u^b-\delta \Gamma_{ab}^k
(\epsilon-\frac{H^2}{2})_{,k} u^a
u^b\nonumber\\&+&(\epsilon-\frac{H^2}{2})_{;ab} (u^a\delta
u^b+u^b\delta u^a)+(\delta\epsilon-H^c\delta H_{c})_{;ab}u^a
u^b\nonumber\\ &=&\delta g^{ab} p_{,ab}^{*}+(u^a\delta u^b+u^b \delta
u^a)p_{ab}^{*}+ h^{ab}(\delta p_{,ab}^{*}-\delta \Gamma_{ab}^k
p_{,k}^{*})\nonumber\\ &+&2\delta [(H^2)_{,c}
\Theta+H^2\Theta_{,c}]u^c\nonumber\\&-&H^b[2\delta(H_{,ab}^a)+
H^k\delta(\Gamma_{ak,b}^a)+2\delta \Gamma_{ak}^a H_{,b}^k]\nonumber\\
&-&2H_{,ab}^a\delta H^b-2\delta \Gamma_{al}^a H^l H_{,b}^b- \delta
(H_{,b}^a) H_{,a}^b-H_{,b}^a\delta (H_{,a}^b)\nonumber\\ &-&\delta
\Gamma_{bl}^a H^l H_{,a}^b-H^aH^c\delta \Gamma_{ac,b}^b- \delta
\Gamma_{ac}^b H_{,b}^c H^a\nonumber\\&+&2x [\sigma \delta \sigma - u^c
u^d(\delta u_{,c}^a u_{a,d}+u_{,c}^a\delta u_{a,d}\nonumber\\
&+&\delta \Gamma_{cl}^a u^l u_{a,d}-u_{,c}^a u_m\delta \Gamma_{ad}^m)-
(u^d\delta u^c+u^c\delta u^d)u_{,c}^a u_{a,d}]\nonumber\\
&+&\frac{1}{2}(\epsilon+3p+H^2)(\delta \epsilon+\delta p+ \delta
H^2)+\frac{1}{2}x(\delta \epsilon+3\delta p+ \delta H^2)\nonumber\\
&+&[2\delta u^c u_{a,c}+2u^c(\delta u_{a,c}- \delta \Gamma_{ac}^m u_m)
](H^a\delta H_b+H^b\delta H^a)\nonumber\\&+& (H^2)_{;a}(\delta
u^a)_{;c}u^c\nonumber\\ &+&(H^2)_{,a}u_{,c}^a\delta u^c+(H^2)_{,a}
u^c(\delta u_{,c}^a+ \delta \Gamma_{cm}^a u_m)
\end{eqnarray}
The perturbed Maxwell's equations are

\beq \delta H_{,0}^b+\delta \Gamma_{0l}H^l=-\delta \sigma_{a}^b
H^a+\frac{2}{3}H^b\delta \Theta-\frac{u^b}{\epsilon+p}\delta p_{,c}
H^c \eeq and from Eq.(3) we find the perturbed equations of motion

\begin{eqnarray}
&-&(\epsilon+p+H^2)\delta u_{,0}^c=h_a^c [H^b\delta H_{,b}^a+\delta
\Gamma_{bl}^a H^b H^l\nonumber\\ &+&H^a\delta H_{,b}^b+\delta
\Gamma_{bm}^b H^m H^a]-\delta h^{cb} p_{,b}^{*}- h^{cb}\delta
p_{,b}^{*}
\end{eqnarray}

where $p^{*}=p+\frac{H^2}{2}$. The non-zero Christoffel symbols
  $\Gamma_{\mu \nu}^{\alpha}$, for the metric (4), are estimated
  easily $\Gamma_{11}^{0}=-\Gamma_{22}^{0}=-\frac{h_{,0}}{2}$,
  $\Gamma_{01}^{0}=\Gamma_{10}^{0}=\frac{h_{,0}}{2(1+h)}$,
  $\Gamma_{13}^{1}=\Gamma_{31}^{1}=\frac{h_{,3}}{2(1+h)}$,
  $\Gamma_{02}^{2}= \Gamma_{20}^{2}=-\frac{h_{,0}}{2(1-h)}$,
  $\Gamma_{23}^{2} = \Gamma_{32}^{2}=-\frac{h_{,3}}{2(1-h)}$,
  $\Gamma_{11}^{3}=\Gamma_{22}^{3}=\frac{h_{,3}}{2}$. We assume that
  $u^{a}=(-1,0,0,0)$, $ \rho=\rho_0$=const.,
  $H^{a}=(0,H^1,0,H^3)$=const., $\delta H^{a}=(0,\delta H^1, \delta
  H^2, 0)$, $\delta u^{a}=(\delta u^0,\delta u^1,\delta u^2,\delta
  u^3)$ and $k^{a}=(0,0,0,k)$ and a plane wave behaviour for the
  perturbed quantities $ \delta \epsilon, \delta p, \delta u^{a}$ and
  $\delta H^a \sim e^{i(kz-nt)}$ and the GW $h\sim e^{i(k_g
  z-n_gt)}$. These simplifications yield a general dispersion relation
  for the coupling of GW with MHD waves.

\begin{eqnarray}
& &\delta \epsilon\{D_0(-n^2-2n^2u_{A}^2)+(kc_s)^2
D_0\nonumber\\&-&
\frac{1}{2}(1+c_s^2)(\epsilon+3p+H^2)D_0\nonumber\\
&-&\frac{1}{2}(1+3c_s^2)(\epsilon+p+H^2)D_0\nonumber\\&+&[n^2+k^2-2n^2u_{A}^2-
2(\epsilon+2p+H^2)](\lambda_1 H^1+\lambda_2 H^1c_s^2)\nonumber\\
&+&[n^2-k^2-2n^2u_{A}^2-2(\epsilon+2p+H^2)](\mu_1 H^3+
\mu_2 H^3c_s^2)\}\nonumber\\
&=&h\{D_0(H^1)^2(\epsilon+2p+H^2-k_g^2)-\lambda_3 H^1-\mu_3 H^3\}
\end{eqnarray}
where

\[
\delta H^1=\frac{1}{D_0}(\lambda_1 \delta
\epsilon+\lambda_2\delta p+\lambda_3 h)\]
\[
\delta H^3=\frac{1}{D_0}(\mu_1 \delta \epsilon+\mu_2\delta p+\mu_3 h)
\]
with \[ \lambda_1=-\frac{H^1}{x}(1-\frac{k^2u_{A}^2}{2n^2}),
~~\lambda_2=\frac{(H^1
H^3)}{x}\frac{H^3}{x}\frac{k^2}{n^2}(1+\frac{k^2}{2n^2}),\]

\begin{eqnarray*}
\lambda_3&=&(-\lambda_1+\lambda_2)[\frac{k^2 (H^3)^2}{2n^2}-
\frac{n_g
x}{n}]\\&+&\frac{(H^1)}{x}\frac{(H^1H^3)^2}{x}\frac{kk_g}{2n^2}
(1+\frac{k^2}{2n^2}),\end{eqnarray*}

\[ \mu_1=-\frac{H^3}{x}(1-\frac{k^2u_{A}^2}{2n^2}),\]
\[\mu_2=\frac{H^3}{x}\frac{k^2}{n^2}[1-\frac{(H^1)^2}{x}-
\frac{k^2}{2n^2}\frac{(H^3)^2}{x}],\]

\begin{eqnarray*}
\mu_3&=&\frac{(H^1 H^3)}{x}\frac{kk_g}{2n^2} H^1[1-
\frac{(H^1)^2}{x}]-\frac{k^3k_g}{4n^4}\frac{(H^1H^3)^2}{x^2} H^3\\
&-&\frac{H^1}{x}\frac{(H^1 H^3)}{x}[\frac{k^2 (H^3)^2}{2n^2}-
\frac{n_g x}{n}](1-\frac{k^2}{2n^2}),
\end{eqnarray*}
and

\begin{eqnarray*}
-D_0&=&1-u_{A}^2-\frac{k^2 u_{A}^2}{2n^2}\\&-&\frac{k^2}{n^2}\frac{(H^3)^2}{x}
\left[1-\frac{k^2}{2n^2}\frac{(H^3)^2}{x}-\frac{(H^1)^2}{x}\right]\\
&+&\frac{k^2 u_{A}^2}{2n^2}\frac{(H^1)^2}{x}+\frac{k^4 u_{A}^2}
{2n^4}\frac{(H^3)^2}{x}.
\end{eqnarray*}
Above, $u_A^2=v_A^2/(1+v_A^2)$ is a relativistic
generalization of the classical Alfv{\'e}n speed 
$v_A=H^2/(4\pi \epsilon/c^2)$ and $c_s=\sqrt{\delta p / \delta \epsilon }$ 
is the speed of sound.
 In passing we note that the non-relativistic Alfv{\'e}n speed $v_A$
generalizes to the relativistic Alfv{\'e}n speed $u_A$ if one does 
not neglect the displacement current in Maxwell's equation, in MHD
theory.

\section{Magnetosonic Waves}
We are interested in investigating the dispersion relation for waves
that correspond to the magnetosonic mode ($H^1\neq 0, H^3=0).$ This is
motivated by the fact that the coupling of MHD waves with GWs is
proportional to $(H^1)^2$. Thus, gravitational waves in the
$z$-direction cannot couple to pure Alfv{\'e}n waves ($H^1=0$, $H^3
\neq 0$). The general relations between the perturbed  quantities,
derived above, simplify when we assume that $H^3=0$  (setting $H^1=H$
from now on)

\begin{eqnarray}
-n^2\delta \epsilon&-&2n^2\frac{H^2}{x}\delta \epsilon + k^2\delta
p-\frac{1}{2}(\delta \epsilon+\delta p)(\epsilon+3p+H^2)\nonumber\\
&-&\frac{1}{2}(\delta \epsilon+3\delta p)(\epsilon+p+H^2)+ (H \delta
H^)[n^2+k^2\nonumber\\ &-&
2n^2\frac{H^2}{x}-2(\epsilon+2p+H^2)]\nonumber\\&=&h(H^2)[(\epsilon+2p+H^2)-k_g^2]
\end{eqnarray}
where, from Maxwell's equations (6) we obtain  \beq (H \delta
H)=\frac{H^2}{1-u_A^2}[\frac{\delta \epsilon}{x}-h \frac{n_g}{n}].
\eeq  From here on, we will write our equations in ordinary units
again.  We combine Eqs.(9)-(10)  to obtain

\begin{eqnarray}
& &\delta \epsilon \{-n^2+(kc_s)^2+k^2
u_{A}^2-(kc_s)^2\frac{u_{A}^2}{c^2}\nonumber\\&-& 2G(4\pi
\rho+\frac{H^2}{c^2})(1+2\frac{c_s^2}{c^2})(1-\frac{u_{A}^2}{c^2})\nonumber\\
&-&4G\frac{u_{A}^2}{c^2}(4\pi \rho+\frac{H^2}{c^2})\}\nonumber\\
&=&h\frac{H^2}{4\pi}\{\frac{n_g}{n}[n^2(1-2\frac{u_{A}^2}{c^2})
+ c^2 k^2-4G(4\pi
\rho+\frac{H^2}{c^2})]\nonumber\\ &+&2(1-\frac{u_{A}^2}{c^2})(4\pi G
\rho+\frac{G H^2}{c^2}-k_g^2c^2)\}
\end{eqnarray}
where $H^2=H_a H^a=H_1H^1$ and we have set $p = 0$ and $\rho=\epsilon/c^2$.

We retain the dominant terms that yield a pure magnetosonic wave in a
strongly magnetized, cold and  tenuous plasma ($c_s<<u_A$) in the
Newtonian limit. This is equivalent to setting $1/c^2 \to 0$ (except
for the $u_A^2/c^2$ terms) and $G\rho \to 0$, $c_s \to 0$ and one
obtains
\begin{eqnarray} (-n^2+k^2u_A^2) \delta \epsilon &=&
h \frac{H^2}{4\pi}  \Biggl\{ \frac{n_g}{n} \left [ n^2 \left (1-2
\frac{u_A^2}{c^2}\right) \right ]  \nonumber\\&+& k^2c^2- 2\left (
1-\frac{u_A^2}{c^2} \right) k_g^2c^2 \Biggr\}. \label{dis}
\end{eqnarray}
In the absence of a GW ($h=0$), the dispersion relation (\ref{dis})
becomes

\beq n_r^2 \simeq k^2u_A^2.\eeq  If one assumes that the frequency of
the driving GW coincides with the frequency of the produced MSW, i.e.
$n=n_g=k_g c$, then the dispersion relation becomes
\begin{equation} (-n_g^2+k^2u_A^2) \delta \epsilon = h \frac{H^2}{4\pi} 
(-n_g^2+k^2c^2), \label{dis2}
\end{equation}
or
\begin{equation}
\left( \frac{\partial^2 }{\partial t^2}-u_A^2\frac{\partial^2
}{\partial z^2} \right) \delta \epsilon=h \frac{H^2}{4\pi} (-n_g^2+k^2c^2),
\end{equation}
i.e. the GW acts as an external parametric oscillator, driving the
MSW. Notice also that, in the absence of a plasma one obtains 
the dispersion relation for GWs propagating in the vacuum.
The two wave numbers differ by an amount $\Delta k$, i.e.
$k=k_g+ \Delta k$, which, in general, is complex, $\Delta k = \Delta
k_R +i \Delta k_I$.  If we denote
\begin{equation}
R= \left(\frac{h}{\delta \epsilon}\right)_0 \frac{H^2}{4\pi},
\end{equation}
(where a subscript 0 denotes amplitude), then (\ref{dis2}) yields a
system of two equations, for $\Delta k_R$ and $\Delta k_I$. Dividing
the two resulting equations, one obtains
\begin{equation}
\frac{-n_g^2 + \left[ k_g^2+2k_g \Delta k_R + \Delta  k_R^2 \right ]
u_A^2} {2k_g \Delta k_I u_A^2} = \frac{A \cos \phi + B \sin \phi }{A
\sin \phi + B \cos \phi},
\end{equation}
where
\begin{equation}
A=\left[ 2k_g \Delta k_R + \Delta k_R^2 \right] c^2,
\end{equation}
\begin{equation}
B=2k_g \Delta k_I c^2.
\end{equation}
and
\begin{equation}
\phi = \Delta k_R z.
\end{equation}
When $|\Delta k_I| < <  | \Delta k_R| \Rightarrow |B/A| < < 1$ and
since $u_A \simeq c$ in a strongly magnetized, tenuous plasma, we
arrive at an expression for $\Delta k_I$
\begin{equation}
\Delta k_I \simeq - \frac{3 k_g}{2} \left( 1 - \frac{u_A}{c} \right)
\tan \Delta k_R z.
\end{equation}
The real part of the wavenumber difference is approximately
\begin{equation}
\Delta k_R = k - k_g =k_g\left(\frac{c}{u_A}-1\right),
 \label{dkR}
\end{equation}
thus, the two waves can linearly interact coherently over a maximum
distance of
\begin{equation}
L\approx \frac{\lambda_g}{2\pi (1-u_a/c)}
\end{equation}
in terms of the GW wavelength $\lambda_g$.

For the plasma around a magnetized compact object, $B^2/ \epsilon \sim
10^2 -10^4$ (where $\epsilon$ is dominated by the kinetic energy of
the plasma particles).  For such a plasma, the interaction length can
be up to several thousand times larger than $\lambda_g$, A more
stringent limit on the interaction length will probably be set but
factors such as the non-uniformity of the magnetic field and the
plasma density with distance from the source.

Assuming also a small imaginary part for the frequency and
substituting $n=n_r + i \gamma$ in Eq.(14), developing $-n^2+k^2 u_A^2
= (n-n_r+i\gamma)(-2n_r)$, and using the Plemelj formula
($\frac{lim}{\gamma \to 0^+}(x+i\gamma)^{-1}= P(1/x)-i\pi \delta(x)$)
we finally arrive at the spatial dependence of the amplitude of the
energy density 
\beq \delta \epsilon_0 \approx \frac{i}{4}  \delta(n-n_g)
h_0 H^2 c \Delta k e^{-i (\Delta k) z}, \eeq  
where $\delta
\epsilon=\delta \epsilon_0 e^{i(nt-kz)}$, $\delta H=\delta H_0
e^{i(nt-kz)}$ and  $ h =h_0 e^{i(n_gt-k_g z)}.$

The magnetic field amplitude is estimated with the use of the
perturbed Maxwell's equations (6) and with the aid of Eq.(10)  
\beq
\delta H_0 \approx \frac{i}{2}  \delta (n-n_g) h_0H c \Delta k e^{-i \Delta
k z}. \eeq 
We note that, the amplitude of the fast magnetosonic wave
is proportional to the GW amplitude. In a tenuous and hot plasma, $u_A
\to c$, so that the mismatch $\Delta k << k$ and our treatment
applies. As a result the amplitude of the MSW grows after having
traveled a distance equal to a half beat between the wave numbers, and
would decline again further out in the absence of dissipation. In a
relativistic electron positron plasma the latter is, however, unlikely
and some observational effects can be expected.

\section{Discussion}

\subsection{Interpretation}

We have found that, to first order, a GW, propagating along the
magnetic field in a uniform and ideal plasma, does not couple with an
Alfv{\'e}n wave (AW) whereas a GW, propagating across the magnetic
field, does couple with a MSW. This can be easily be understood as follows.

\begin{figure}
\resizebox{\hsize}{!}{\includegraphics{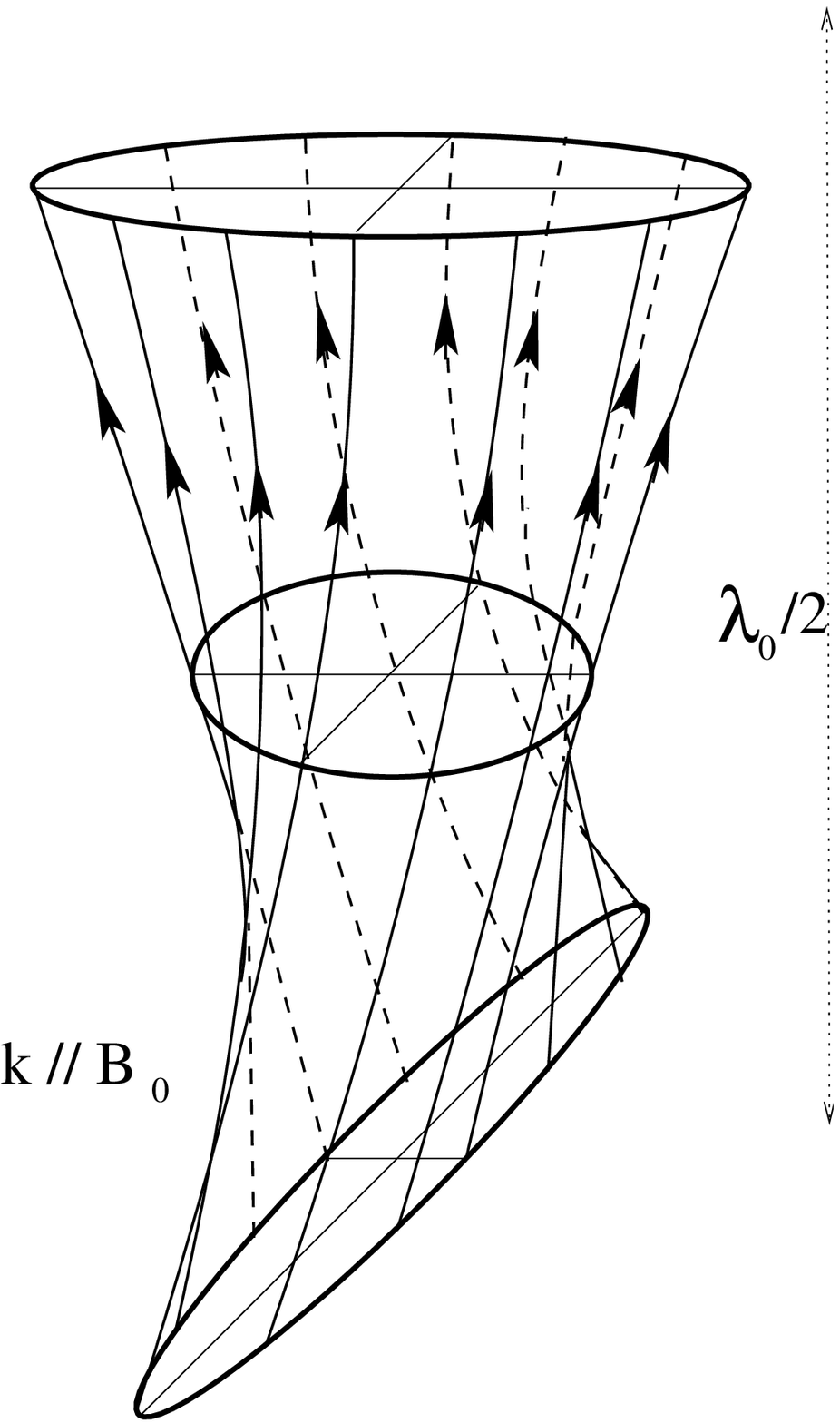}}

\caption[]{Sketch of the effect of a large-amplitude GW traveling
  parallel to a uniform magnetic field frozen into an ambient ideal
  plasma, $\vec k \parallel \vec{B_0}$ at an arbitrary time. Here the
  vertical direction coincides with the direction of wave propagation
  and of the background magnetic field.}
 \label{f1}
\end{figure}

In Fig. \ref{f1} we have sketched the effect of a large-amplitude GW
traveling {\it along} an, initially uniform, magnetic field embedded
in an ideal plasma. At first sight it is surprising that the GW would
not be able to excite an AW as the perturbed magnetic field shows
clear distortions which in an ordinary magnetoplasma would be expected
to generate AWs. However, the point is that the energy of an AW with
amplitude $\delta H^1$ is proportional to $(\delta H^1)^2$ whereas, for a MSW,
the magnetic part of the wave energy density is proportional to $ H^1
\delta H^1$. As a result, the energy in the AW is of higher order in the
perturbation as compared to the magnetosonic case. Therefore, our
first order calculation should not lead to the excitation of AWs, as
indeed we have found. On the other hand, we do expect such a coupling
to show up in a higher-order calculation, and such a coupling will be
relevant in case the GW is non-linear.

\begin{figure}

\resizebox{\hsize}{!}{\includegraphics{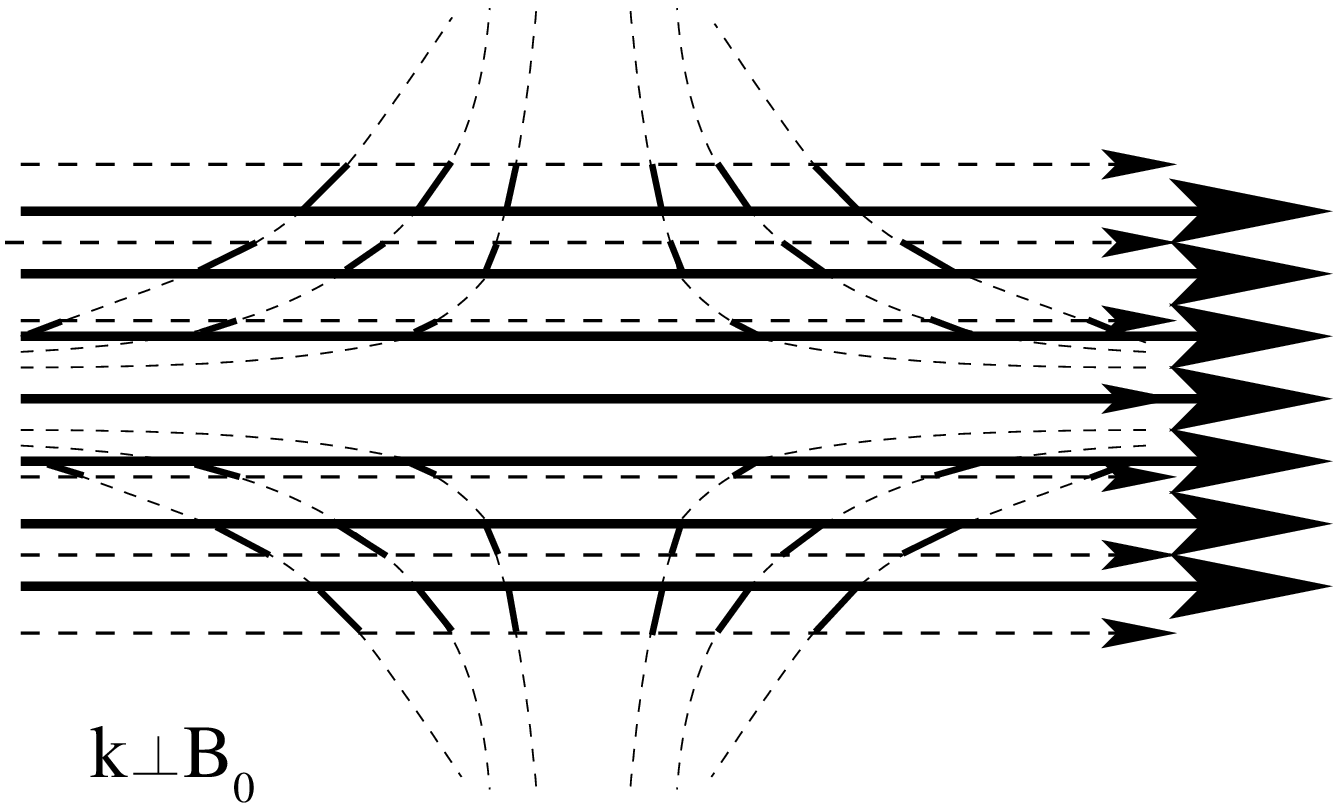}}

\caption[]{Sketch of the effect of a GW traveling perpendicular
to a
  uniform magnetic field frozen into an ideal plasma. The tidal motion
  of the GW is indicated with curved dashed lines. The drawn and,
  respectively, dashed arrowed lines are the magnetic field lines half
  a cycle apart. Their individual relative displacements are shown
  by the bold sections along the tidal curves. Note that the field
  lines remain straight under the action of the GW.}
 \label{f2}

\end{figure}

In Fig. \ref{f2} we have sketched the effect of a GW traveling at
right angles to an initially uniform magnetic field frozen into an
ideal plasma. Clearly, the field is periodically compressed by the GW
although the differential motions induced by the GW are
divergence-free! At the same time, the ambient uniform magnetic field
remains uniform and straight and does not become bent. As a result, no
AW excitation is expected to occur by the GW even to higher order.

Further, our result that the coupling between a GW and MSWs vanishes
as soon as the dispersion relations for both wave modes coincide
exactly, can be understood as follows. An effective transfer of energy
from one wave mode to the other either requires that there is some
dissipation (Weber 1960; Esposito 1971) or, alternatively, that the
signal speeds of both modes differ. If the plasma is both ideal and
the group speeds of the wave modes coincide, there is no time lag
during which some of the perturbed kinetic/magnetic plasma energy
established by the GW can be fed into a magnetosonic wave. In a very
general sense, the changes brought about by the GW in the
magnetoplasma are completely adiabatic and the periodic changes in the
plasma completely reversible. All free energy in the magnetic field
reverts back into the GW as it leaves the plasma again.

\subsection{Applications and perspective}

A number of papers exist which discuss the interaction of MSWs with a plasma,
e.g.  coherent MSW-particle interactions, cascading of long wave-length
MSWs to short wavelengths and subsequent particle acceleration from
turbulent diffusion (see Dermer at al. 1996), and finally the
interaction of a relativistic magnetosonic shock wave with a
positron-electron plasma (Hoshino et al. 1992). Thus, if a
magnetosonic wave is driven by a GW, then it has the
potential of producing additional interesting phenomena. A more
detailed understanding of the linear and nonlinear evolution of the
excited MSWs under the conditions of the particular GW source is
needed.  Clearly, it is beyond the scope of the present study to model
the acceleration and radiation properties of a GW-driven MSW in the
vicinity of a GW-source but we want to mention briefly two specific
applications which have been the motivation for us to undertake this
study: GWs excited by star quakes or rotational instabilities in
neutron stars (see Andersson, Kokkotas \& Schutz 1999) or magnetars,
and GWs excited in merging neutron star binaries and their possible
effects on gamma ray bursts (GRBs). 

In particular, we expect that the driving of MSW by gravitational
waves can become interesting near strongly-magnetized compact sources
of high-frequency GWs. The non-radial modes of
relativistic stars produce GWs with frequencies of the
order of kHz.  These can be excited in several ways: in the 
core-collapse leading to the formation of a neutron star, several modes
of pulsation can be excited to large amplitudes. In newly-born
neutron stars the quadrupole $f$- and $r$-modes can become unstable to
the emission of gravitational radiation, due to the CFS instability
(see Stergioulas 1998, Andersson \& Kokkotas 2001 for recent
reviews). In the merger of two neutron stars, if a black hole is not
formed promptly, the resulting merged object will emit large amplitude
GWs, due to the excitation of it's nonradial modes
(see Shibata and Uryu, 2001). If strongly magnetized neutron stars
(magnetars) exists, then the coupling of their secular spin-down
with their very large magnetic field can lead to crust-fracturing and
excitation of the quadrupole torsional mode in the crust of the star 
(see Duncan 1998). In all the above cases, high-frequency gravitational
waves travel through the magnetosphere of the compact object.

In addition to GWs emitted by normal-mode excitation in
neutron stars, the main frequency of GWs produced in
the late stages of a neutron star binary inspiral is several hundred
Hz to a few kHz (see e.g. Duez, Baumgarte and Shapiro 2000). In such a
case, however, the geometry of the magnetic field changes rapidly and this
has to be taken into account.

\section{Summary and Conclusion}

In this article we study for the first time the coupling of
magnetosonic waves with GWs. Using the linearized
general-relativistic hydromagnetic equations, we found the general
dispersion relation for coupling of GW with the MHD waves.  We showed
that in the linear approximation only the fast magnetosonic waves will
be coupled with the GW.  We analyzed the linear dispersion relation
for the GW-MSW coupling, and found the relation between the amplitudes
of the two waves, when they are in resonance.

In carrying out this study we have made several simplification: (1) we
treat only one polarization for the GW (see Eq. (4)), (2) the GW was
treated as an external driver for the plasma, (3) we ignore non-linear
terms and obliquely propagating waves. The above points, as well as
the growth rate of driven MSW waves for specific GW sources, will be
discussed in future work.

\begin{acknowledgements}
We thank Dr. K. Kokkotas for making several suggestion and
keeping up with our work and Bernard Schutz, Joachim
Moortgat and Gerard 't Hooft for stimulating discussions.
NS acknowledges the support by the European Union grant
HPRN-CT-2000-00137.

\end{acknowledgements}

\end{document}